\documentclass[12pt]{article}
\usepackage{epsfig}
\usepackage{amsfonts}
\usepackage{amsopn}
\usepackage{amsmath}

\title
{\vskip -50 pt
\begin{flushright}
\normalsize\rm  NORDITA 2026-002
\end{flushright}
\vskip 20 pt
Tensionless spinning string: emergence of world-sheet torsion and  covariant density of energy-momentum tensor 
}

\author{
 A. A. Zheltukhin 
 \thanks{E-mail: aaz@nordita.org}  \\
Kharkov Institute of Physics and Technology, \\
1, Akademicheskaya St., Kharkov, 61108, Ukraine \\  
NORDITA, KTH Royal Institute of Technology and \\Stockholm University,
Hannes Alfv\'{e}ns v\"{a}g 12, \\
SE-106 91 Stockholm, Sweden 
}

\date{}
\begin{document}
\maketitle

\begin{abstract}

The action of tensionless spinning string invariant under reparametrizions, both  local supersymmetry and dilatations, is considered. The density of energy-momentum tensor is constructed and vanishing of  its covariant divergence is proved.  This result arises from mutual cancellation of the bosonic and fermionic contributions. Differences in the geometry of worldsheets swept by tensionless and tensionfull spinning strings are analyzed. Shown is emergence of covariant trace of a torsion tensor on w-s of the tensionless spinning string. It is derived from the condition for the fermionic scalar density to be a composite one including the 2-dim. w-s density simulating the 4-dim. Rarita-Schwinger field. The said condition is accompanied with the Noether condition for covariant divergence of the vector metric density to vanish.

\end{abstract}

\section{Introduction}

Going beyond the Standard Model requires new ideas for better understanding of  space-time structure, quantum field theory and astroparticle physics at the Planck scale, where the coupling constants of all fundamental interactions become equal.  
Advanced analysis of the high energy symmetries of scattering amplitudes in string, in particular was made in \cite{GrMe, Gros} with  hope of discovering an enormous spontaneously broken symmetry. This symmetry could be hidden in complex nonlinear structure  of the interaction amplitudes. 
This structure was simplified in perurbative calculations in the limit of very high energies,  when the Planck mass $M_{p}$, becomes negligible. This limit is equivalent to the zero tension $T=\frac{1}{\pi\alpha'}$ in string theory, as $T\sim M_{p}$. 
 In this scale invariant limit all particles have  zero mass,  and full unbroken symmetry group of string theory is restored.  Convincing arguments have been presented by the authors,  which show that this group is an infinite parameter one. This allowed to express scattering amplitudes of all string states through the dilaton scattering amplitudes\footnote{It is interesting to note an important role of scale symmetry in analysis of the  experimental data \cite{Planck}.}.

On the other hand, interest in studying the dynamics of strings at the limiting point $\alpha'=\infty$ arose after the work \cite{Shil}. In the  latter, the w-s of the tensionless string, called null (lightlike) string, was treated as a 2-dim. null geodesic surface generalizing 1-dim. world line of massless particle. 
In particular, studied in the literature were various physical and mathematical aspects of the classical and quantum dynamics of null (super)strings, null spinning strings, their interactions with external fields, as well as generalization of the supersymmetric string action formulations to the case of (super)-p-branes (see e.g.  \cite{VZ}--\cite{DimU}).  It has been shown that many models of supersymmetric extended zero objects have no critical dimension and can be covariantly quantized using well-known BRST/BFV formalisms.  However, critical dimensions, depending on the number of space-time global supersymmetries N, appear in models of the tensionless conformal bosonic
 \cite {GLSSU_cfns} and spinning \cite{Salt_susy} strings. The Hamiltonian BRST  quantization of  these models led to the critical dimension D=2 for N=0 and negative dimensions at N  ${\neq 0}$, respectively. This points to the role of space-time symmetries and geometries of target spaces in which tensile and tensionless extended objects move.  A special role of target spaces with a possibly degenerate metric is discussed in \cite{BorstGalKim} along  with review of progress in research  of  symmetries of  tensionless string  and branes. 
All that stimulates study of hypersurfaces (worldvolumes) of supersymmetric models of tensionless  strings, $p$- branes and $(p+1)$-dim. physics on them.

An interesting example of 2-dim. w-s geometry created by the tensionful  spinning string with action invariant under local supersymmetry and reparametrizations, was constructed  in \cite{BdVH}, \cite{DeZu}.  As revealed there, this action generated a two-dimensional model that described the interaction of matter and supergravity. This model contains the string tension, and it is promising  to analyze changes in the worldsheet  geometry  in the limit of zero tension. For the case of bosonic relativistic null string the metric on its w-s becomes degenerate  (i.e. $det(g_{\mu\nu}) =0$) at the limiting point $T=0$. It caused use of a  w-s vector density $\rho^{\mu}$ instead of the w-s metric tensor $g_{\mu\nu}$. 
Due to the use of $\rho^{\mu}$ in formulations of invariant actions of locally supersymmetric null objects a fermionic counterpart of $\rho^{\mu}$ is forced to turn into a w-s density. The same thing happens with the reduced two-dimensional multiplet of supergravity. As a result, its covariant derivative gets the well-known additional term $\Gamma_{\mu}$ equal to the trace of an affine connection 
$\Gamma^{\beta}{}_{\mu}{}_{\beta}$. 
This feature has been taken into account under reformulation of the invariant action for bosonic null strings to the spinning string action in the limit of zero tension (see  \cite{LST258} and refs. there). The reformulated action invariant under local supersymmetry and w-s diffeomorphisms is suitable for comparison with the invariant action for the tensionful spinning string. 
The results of such a comparison are analyzed here, and differences in the geometry of worldsheets swept by tensionless and tensionfull spinning strings are presented. Taking into account  the presence  of scale symmetry in the  scenario \cite{GrMe, Gros} and in  the  analysis of cosmological data \cite{Planck} this symmetry is added  in the action \cite{LST258}. 
This is achieved by going to minimally extended covariant derivatives, including the w-s gauge field $a_{\mu}$ for the dilatation symmetry.  The use of such extended derivatives leads to an action invariant under the three sought-for local symmetries, where $a_{\mu}$ is treated as an independent auxiliary field. 
The latter naturally appears after conjecture on a composite structure, similar to the  one used  in \cite{ZhCom)},  of the auxiliary scalar density $a(\xi)$ considered as the Lagrange multiplier connected with the space-time dilatation symmetry (see \cite{DimU} and refs. there).

 Variations of this action produce the generalized Euhler-Lagrange equations expressed in terms of covariant blocks constructed from w-s fields, densities and their extended  covariant derivatives.  Using these equations we construct the energy-momentum tensor density  for the tensionless spinning string and prove vanishing of this density and its covariant divergence.
We find that transition to zero tension is accompanied with reduction of the w-s  torsion tensor of the tensionfull spinning string to its covariant trace.  Such a reduction follows from the condition for the primary fermionic scalar density in the action to be a composite one including 2-dim. w-s density as an image of the 4-dim. Rarita-Schwinger field. We show that the said compositness condition demands vanishing of the extended covariant divergence of the above mentioned vector metric density.

\section{Torsion in model of tensionless spinning string}

The relativistic invariant action of N=1 closed spinning string in the n-dimensional Minkowski space with the metric $\eta _{mn}=diag. (1, -1, ..., -1)$  has the following form
\begin{eqnarray}\label{actn}
 S =  \frac{1}{2}\int d^{2}\xi
 \{ (\rho^{\mu}D_{\mu}X^{m} + i\chi\Sigma^{m})
(\rho^{\nu}D_{\nu}X_{m} + i\chi\Sigma_{m})
+i \Sigma^{m}\rho^{\nu}\partial_{\nu} \Sigma_{m} \},
\end{eqnarray}
where  $\xi^{\mu}=(\tau, \sigma)$ are the w-s  parameters, 
$ X^{m}(\xi) $ are the space-time  coordinates of  the spinning string with their fermionic partners  $\Sigma^{m}(\xi)$.  
The field $ \rho^{\mu}(\xi)$ is  a  w-s vector density 
 weight $ \omega_{\rho}=\frac{1}{2}$  and  $ \Sigma^{m}(\xi) , \chi(\xi) $ are the  Grassmann scalar w-s densities with 
$\omega_{\Sigma}=\omega_{\chi}=\frac{1}{4}$.
The  transformation laws of these  densities under  w-s diffeomorphisms
 $\xi^{'\mu}=f(\xi^{ \mu} )$ are 
\begin{eqnarray}\label{dens}
\rho^{'\mu}(\xi')=J^{-\frac{1}{2}}\frac{\partial \xi^{' \mu}}{\partial \xi^{\alpha}}\rho^{\alpha}(\xi) ,  
\,  \,  \, 
\Sigma^{' m}(\xi')=J^{-\frac{1}{4}}\Sigma^{m}(\xi), \,  \,  \chi{'} (\xi')=J^{-\frac{1}{4}}\chi(\xi),  \,  \,  \
 ( d^{2}\xi)^{'}=J(\xi) d^{2}\xi,  
\end{eqnarray}
  where $ J= \det \frac {\partial  \xi^{'\mu}} {\partial \xi^{\alpha}}$ is the w-s Jacobian,  
so that action (\ref{actn}) is  invariant under diff-s. 
$S$ is also invariant under the gauge scale transformations defined by the relations $\xi'^{\mu}=\xi^{\mu}$  and 
\begin{eqnarray}\label{scasy}
X^{'m}(\xi)=e^{\Lambda(\xi)}X^{m}(\xi),   \,  \rho^{'\mu}=e^{-\Lambda}\rho^{\mu}, \,  \, 
\Sigma^{'m}=e^{\frac{1}{2}\Lambda}\Sigma^{m}, \,  \, 
\chi'=e^{-\frac{1}{2}\Lambda}\chi,  \, \,   a^{'}_{\mu}(\xi)= a_{\mu}(\xi) - \partial_{\mu}\Lambda(\xi),
\end{eqnarray}
 for the w-s fields,  where $a_{\mu}(\xi) $ is the w-s vector gauge field for scaling. 
The scale-covariant w-s derivatives  $D_{\mu}$ for the fields are 

\begin{eqnarray}\label{covder}
D_{\mu}X^{m}:= \partial_{\mu}X^{m} + a_{\mu}X^{m}, \, \, 
D_{\mu}\Sigma^{m}:=\partial_{\mu}\Sigma^{m} +\frac{1}{2}a_{\mu}\Sigma^{m}, \nonumber
\\
 D_{\mu}\rho^{\nu}:= \partial_{\mu}\rho^{\nu} - a_{\mu}\rho^{\nu},  \, \, 
D_{\mu}\chi:=\partial_{\mu} \chi - \frac{1}{2}a_{\mu}\chi 
\end{eqnarray}
and  they contain their scale charges $q_{X}=-q_{\rho} =1, 
 \, q_{\Sigma}=-q_{\chi}=\frac{1}{2}$  and $q_{a_{\mu}}=0$.   
Then extended w-s derivatives ${\bf D}_{\mu}$ covariant under all local symmetries take the following form  
\begin{eqnarray}\label{totcovd}
{\bf D}_{\mu}X^{m}= D_{\mu}X^{m},  \, \,  \, \,  {\bf D}_{\mu}\rho^{\nu}= D_{\mu}\rho^{\nu} +\Gamma^{\nu}{}_{\alpha}{}_{\mu} \rho^{\alpha} - \frac{1}{2}\Gamma^{\beta}{}_{\mu}{}{}_{\beta}\rho^{\nu},
\nonumber  \\
{\bf D}_{\mu}\Sigma^{m}=D_{\mu}\Sigma^{m} 
-  \frac{1}{4}\Gamma^{\beta}{}_{\mu}{}_{\beta}\Sigma^{m},   \, \, 
 {\bf D}_{\mu}\chi=D_{\mu} \chi - \frac{1}{4}\Gamma^{\beta}{}_{\mu}{}_{\beta}\chi, 
\, \,  {\bf D}_{\mu}a_{\nu}=\nabla_{\mu}a_{\nu}=
\partial_{\mu}a_{\nu} - \Gamma^{\alpha}{}_{\nu\mu}a_{\alpha}.
\end{eqnarray}
We notice the presence of the additional term $(-w\Gamma^{\beta}{}_{\mu}{}_{\beta})$ in the covariant derivatives of the tensor densities  in (\ref{totcovd}). 
Note that $\nabla_{\mu}$ denotes derivatives 
covariant under w-s diffeomorphisms   
and their use allows to represent (\ref{totcovd}) in the condenced form 
\begin{eqnarray}\label{nab}
{\bf D}_{\mu}X^{m}=  \partial_{\mu}X^{m} + a_{\mu}X^{m},    \, \,  \,      
 {\bf D}_{\mu}\rho^{\nu}= \nabla_{\mu}\rho^{\nu} - a_{\mu}\rho^{\nu}
\nonumber   \\
{\bf D}_{\mu}\Sigma^{m}=\nabla_{\mu}\Sigma^{m}  + \frac{1}{2}a_{\mu}\Sigma^{m},  \, \,  \, 
{\bf D}_{\mu}\chi=\nabla_{\mu} \chi - \frac{1}{2} a_{\mu} \chi. 
\end{eqnarray} 
Then $S$ is presented in the explicitly invariant form under diffemorphisms and dilatations  
 \begin{eqnarray}\label{actneq}
 S =  \frac{1}{2}\int d^{2}\xi
 \{ (\rho^{\mu}{\bf D}_{\mu}X^{m} + i\chi\Sigma^{m})
(\rho^{\nu}{\bf D}_{\nu}X_{m} + i\chi\Sigma_{m})
+i \Sigma^{m}\rho^{\nu}{\bf D}_{\nu} \Sigma_{m} \}.					\end{eqnarray}
 The connection coefficients $ \Gamma^{\alpha}{}_{\nu\mu}$ in (\ref{actneq}) localized in the kinetic term $ \Sigma^{m}\rho^{\nu}{\bf D}_{\nu} \Sigma_{m} $ (\ref{actneq}) are free due to the relation
 $\Sigma^{m}\Sigma_{m} \equiv 0$ defining the Grassman numbers. 
Then $\Gamma^{\alpha}{}_{\nu\mu}$ can be viewed as components of an affine connection whose subindices are not symmetric under permutations.

The action $S$ (\ref{actn}) is also invariant under the local supersymmetry transformations 
 \begin{eqnarray} \label{lsusY}
\delta X^{m}=- \varepsilon\Sigma ^{m}, \, \, \, \, \, \, 
\delta \Sigma^{m}=-i(\rho^{\mu}D_{\mu} X^{m} 
+\frac{i}{2}\chi\Sigma^{m}) \varepsilon, \nonumber  
\\
\delta\rho^{\mu}=\varepsilon\chi\rho^{\mu}, \, \, \, 
\delta\chi=-i\rho^{\mu}(\partial_{\mu} + a_{\mu})\varepsilon 
+  \frac{i}{2} \varepsilon\partial_{\mu}\rho^{\mu}, \, \, \, 
\delta a_{\mu}= 0.  
\end{eqnarray}
  The infinitesimal parameter $\varepsilon$ in (\ref{lsusY}) is a scalar density with $\omega_{\varepsilon}=-\omega_{\chi}=-\frac{1}{4}$ and $q_{\varepsilon}=-q_{\chi}=\frac{1}{2}$,  so that the w-s derivative of $\varepsilon$ takes the form ${\bf D}_{\mu}\varepsilon$
 \begin{eqnarray} \label{dervare}
D_{\mu}\varepsilon:=\partial_{\mu} \varepsilon + \frac{1}{2}a_{\mu}\varepsilon  \, \, 
\rightarrow  \, \,
 {\bf D}_{\mu}\varepsilon=D_{\mu} \varepsilon + \frac{1}{4}\Gamma^{\beta}{}_{\mu}{}_{\beta}\varepsilon.
\end{eqnarray}
 covariant under reparametrizations, local scale and supersymmetry transformations.   
Relations  (\ref{totcovd}) allow to present the variation $\delta\chi$ from  $(\ref{lsusY})$ in the explicitly covariant form
\begin{eqnarray}\label{varchi}
\delta\chi=-i\rho^{\mu} {\bf D}_{\mu}\varepsilon 
+ \frac{ i}{2} \varepsilon{\bf D}_{\mu}\rho^{\mu}.
\end{eqnarray}

As emphasized above, the w-s fermion fields in (\ref{actn}) do not have their own spinor or vector indices usually used to describe gravitino in 4-dim. space-time. For this reason, the w-s scalar density $\chi$ has transformation properties (\ref{varchi}) that differ from those associated with gauge fields.  
This discrepancy can be overcome by assuming that the field density $\chi$ has a composite structure. The latter may be treated as formed by a hypotetic bound state of w-s
 vector densities with opposite statistics. Then the merger of these vector fields can be mathematically desribed by the construction of the w-s scalar condensate:  $\chi=\rho^{\mu}\chi_{\mu}$.

The substitution of this product into (\ref{varchi}) transforms it into the equation for $\delta\chi_{\mu}$ 
\begin{eqnarray}\label{varchim}
\chi=\rho^{\mu}\chi_{\mu} 
\, \,  \rightarrow \, \, \rho^{\mu}(\delta\chi_{\mu}+ i{\bf D}_{\mu}\varepsilon)
=  \frac{ i}{2} \varepsilon{\bf D}_{\mu}\rho^{\mu}.
\end{eqnarray}
The requirement for $\chi_{\mu}$ to be the gauge ﬁeld of the w-s susy  
\begin{eqnarray}\label{chiml} 
\delta\chi_{\mu}=-i{\bf D}_{\mu}\varepsilon, 
\end{eqnarray}
is achieved if the covariant condition for $\rho^{\mu}$ is imposed
 \begin{eqnarray}\label{conchi}
 {\bf D}_{\mu}\rho^{\mu}=0.
\end{eqnarray}  
This condition is a consequence of the second Noether theorem which manifests linear dependence of equations of motion in theories invariant under local symmetries. Eq. (\ref{conchi}) yields such a dependence between the componets of $\rho^{\mu}$ and the variational equations of motion, respectively, caused by the emergence of the gauge density $\chi_{\mu}$ in  (\ref{actn}). Using the product $\chi=\rho^{\mu}\chi_{\mu}$ in  representation (\ref{actneq}) of $S$ transforms it to the form  
\begin{eqnarray}\label{comp_actn}
 S =  \frac{1}{2}\int d^{2}\xi 
 \{\rho^{\mu} \rho^{\nu} ({\bf D}_{\mu}X^{m} +
 i\chi_{\mu}\Sigma^{m})
({\bf D}_{\nu}X_{m} + i\chi_{\nu}\Sigma_{m})
+i\Sigma^{m} \rho^{\nu}{\bf D}_{\nu} \Sigma_{m} \},
\end{eqnarray}
which shows that $\rho^{\mu} \rho^{\nu}$ transforms under diffeomorphisms similarly to the multiplier $\sqrt{-g}g^{\mu\nu}$, invariant under the Weyl transformation, in the action of  tensile string. 

For further analysis, it is important to remember that for the case of  tensile spinning string the gauge field $\chi_{\mu}$ of the local supersymmetry with the transformation law  
\begin{eqnarray}\label{chimtnl}
\delta\chi_{\mu}=-i D_{\mu}\alpha
\end{eqnarray}
was a spinor fermionic counterpart to the w-s zweibein $e^{a}_{\mu}$.
At the same time the covariant derivative for the susy parameter
 $\alpha$ took the form 
\begin{eqnarray}\label{chimcvdtnl}
 D_{\mu}\alpha=\partial_{\mu}\alpha - \frac{1}{2}\omega_{\mu}\gamma_{5}\alpha.
\end{eqnarray}
Similarity of transformation laws (\ref{chiml}) and (\ref{chimtnl}) is achieved through different realizations of the local supersymmetry parameters adapted to different representations of the gauge field $\chi_{\mu}$.
In the zero tension llimit the spinor parameter $\alpha(\xi)$ is replaced  by the w-s density 
$\varepsilon(\xi)$. The conjecture about composite structure of $\chi(\xi)$ reveals the hidden role of 
$\chi_{\mu}$ as a fermionic superpartner of $\rho^{\mu}$. 
In this regard, we recall that for the tensile spinning string its w-s multiplet created an image of the 4-dim supergravity multiplet where  the spinor $\chi_{\mu}$ simulated the Rarita-Schwinger field.
 This proves the statement that the pair $(\rho^{\mu}, \chi_{\mu})$ in  $S$ (\ref{actn}) simulates a reduced supergravity multiplet caused by the transition to zero tension, but preserving its coupling with the matter fields and the gauge field $a_{\mu}$ of the scale symmetry.

However, in the case of tensile spinning string its local supersymmetry required the presence of an additional term quadratic in the spinor 
$\chi_{\mu}$. As a result, $D_{\mu}\alpha$ became an extended derivative covariant under reparametrizations, local supersymmetry and Lorentz transformations. It was achieved by using the antisymmetric w-s spin connection $\omega_{\mu ab}$
\begin{eqnarray}\label{spincon}
 \omega_{\mu}: =-\frac{1}{2}\varepsilon^{ab}\omega_{\mu ab},
\end{eqnarray}
with the following spinor structure of the abelian gauge field 
$\omega_{\mu}$ 
\begin{eqnarray}\label{spinconex}
 \omega_{\mu}= \omega_{\mu}^{(0)} +\frac{i}{2}{\bar\chi}_{\mu}\gamma_{5}\gamma^{\rho}\chi_{\rho}. 
\end{eqnarray} 
The first term $\omega_{\mu}^{(0)}$ in (\ref{spinconex}) encoded the Christoffel connection constructed from the zweibein components and their inverses ${e^\mu}_{a}$. The second term in (\ref{spinconex}), being a Lorentz covariant bilinear in the spinor $\chi_{\mu}$, extended the Riemannian connection to an affine connection. 
This extension was classified as  produced by the torsion tensor
\begin{eqnarray}\label{torcon}
 C^{a}_{\mu\nu}:=D_{\mu}e^{a}_{\nu} - D_{\nu}e^{a}_{\mu}=\frac{i}{2}{\bar\chi}_{\mu}\gamma^{a}\chi_{\nu}
\end{eqnarray}
describing the w-s geometry of the embedded tensile spinning string. 

To determine the structure of the torsion tensor, associated with  the  tensionless string, one can compare the explicit form (\ref{dervare}) of the covariant derivative ${\bf D}_{\mu}\varepsilon$
\begin{eqnarray} \label{codervare}
{\bf D}_{\mu}\varepsilon=\partial_{\mu} \varepsilon + \frac{1}{2}(a_{\mu} + \frac{1}{2}\Gamma_{\mu})\varepsilon, \, \, \, \, \,  \Gamma_{\mu}:=\Gamma^{\beta}{}_{\mu\beta}
\end{eqnarray}
 with  (\ref{chimcvdtnl}) using the explicit form of $\omega_{\mu}$  (\ref{spinconex}).
The generalized connection in (\ref{codervare}) contains the abelian gauge field $a_{\mu}$ and the trace
 $\Gamma^{\beta}{}_{\mu\beta}$ which transforms under reparametrizations as 
\begin{eqnarray}\label{abel} 
\Gamma'_{\mu}(\xi')=\frac{\partial \xi^{\beta}}{\partial \xi'^{\mu}}\Gamma_{\beta}(\xi) + 
\frac{\partial \xi'^{\lambda}}{\partial \xi^{\beta}}
\frac{\partial^{2}\xi^{\beta}}{\partial \xi'^{\mu} \partial \xi'^{\lambda}}.
\end{eqnarray}
This law and the scale transformations (\ref{scasy}) of $a_{\mu}$ guarantee complete covariance of ${\bf D}_{\mu}\varepsilon$ under all local symmetries of  (\ref{actn}).
The representation of $\Gamma^{\beta}{}_{\mu\beta}$(\ref{codervare}) in the form
\begin{eqnarray}\label{Gam}
\Gamma_{\mu}=\frac{1}{2}(\Gamma^{\beta}{}_{\mu}{}_{\beta} + \Gamma^{\beta}{}_{\beta}{}_{\mu}) 
+ \frac{1}{2}(\Gamma^{\beta}{}_{\mu}{}_{\beta} - \Gamma^{\beta}{}_{\beta}{}_{\mu}) \equiv
\Gamma^{\beta}{}_{(\mu\beta)} + \frac{1}{2} S_{\mu}     
\end{eqnarray}
 shows that the second term in (\ref{Gam}) is equal to half the torsion tensor  
\begin{eqnarray}\label{Gamtor}
 S^{\alpha}{}_{\mu\beta}:=\Gamma^{\alpha}{}_{\mu}{}_{\beta} - \Gamma^{\alpha}{}_{\beta}{}_{\mu}
\end{eqnarray}
contracted over its upper and lower indices. 
This term, denoted by $S_{\mu}$, is a worldsheet vector.
The first term in (\ref{Gam}) is equal to the symmetric part of the  affine connection which does not coincide with the Christoffel symbols if  
$S_{\mu}\neq 0$. Otherwise, if $S_{\mu}=0$, this term yields  the contracted Christoffel connection
\begin{eqnarray}\label{Riemn}
S_{\mu}=0 \longrightarrow \, \,  \, \,    \frac{1}{2}(\Gamma^{\beta}{}_{\mu}{}_{\beta} + \Gamma^{\beta}{}_{\beta}{}_{\mu}) =\partial_{\mu}ln\sqrt{-g}. 
\end{eqnarray} 
  The full torsion tensor remains incompletly defined, unlike the tensor  of the tensile string, all components of which are given by the bilinear covariant
 $\frac{i}{2}{\bar\chi}_{\mu}\gamma_{5}\gamma^{\rho}\chi_{\rho}$.  
It is explained by the loss of both spinorial structure of nilpotent density 
$\chi_{\mu}$  (\ref{varchim}) and some components of $\Gamma^{\alpha}{}_{\mu}{}_{\beta}$.  
Such peculiarities in the dynamics of tensionless spinning strings are results of using w-s tensor densities in the model (\ref{actn}). 
On the other hand covariant derivatives of tensor densities contain the contracted affine connection $\Gamma_{\mu}(\xi)$ and this creates a protective mechanism against the elimination of torsion from the equation of motion of tensionless strings.  
It can be seen from the equations of motion and the Noether constraint (\ref{conchi})
\begin{eqnarray}\label{coneq}
 {\bf D}_{\mu}\rho^{\mu}\equiv 
D_{\mu} \rho^{\mu}+ \frac{1}{2}\Gamma^{\beta}{}_{\mu}{}_{\beta} \rho^{\mu}
\equiv \partial_{\mu}\rho^{\mu} - (a_{\mu} - \frac{1}{2}\Gamma_{\mu})\rho^{\mu}=0.
\end{eqnarray}

Let us also remember that the Grassmann space-time vector
 $\Sigma^{m}(\tau)$, introduced for the description of Dirac particles with spin 1/2, was a scalar under reparametrizations of their world lines. 
However, in the invariant action (\ref{actn})  $\Sigma^{m}(\xi)$ has turned into a scalar density on the w-s swept by tensionless spinning string. Such a regeneration is explained by a new role of $\Sigma^{m}$ as an collective variable carrying spin degrees of freedom. 
Really, in string picture $\Sigma^{m}(\xi)$ can be interpreted as an order parameter describing the relativistically covariant collectivization of the space-time spin variables of interacting massless spinning particles. 

 From the point of view of 2-dim.  local supersymmetric theory on the string worldsheet, this collectivization manifests itself in the appearance of the fermionic density $\chi_{\mu}(\xi)$ accompanied with $S_{\mu}(\xi)$.  Emergence of this w-s vector in the description of tensionless spinning strings confirms connection between torsion and spin revealed in the model of the tensile spinning string. 
Such a correspondence  between relativistically invariant physics in ambient multidimensional (super)spaces and on hypersurfaces embedded in them may be useful for applications in relativistic systems containing spin degrees of freedom.

Thus,  the phenomena discovered in models of tensionless spinning
 strings and branes may shed new light on hidden connections between  (super)gauge theories and (super)gravity. 

In the next Section we consider equations of motion produced by action (\ref{actn}) and discuss the appearance of the Lagrangian constraints expressed in terms of the covariant densities.  
The correspondence between the Lagrangian and Hamiltonian constraints is discussed.

\section{Equations of motion and the Lagrangian constraints}

To analyse the variational equations and constraints of the model we  use the w-s density $Z^{m}, \,  \omega_{Z}=1/2$  and the vector density $Y^{m \mu},  \, \omega_{Y}=1$ given by 
\begin{eqnarray}\label{Z,Y}
Z^{m}:  =\rho^{\nu}D_{\nu}X^{m} + i\chi\Sigma^{m},   \, \, \,    Y^{m\mu}: = \rho^{\mu}Z^{m}. 
\end{eqnarray}
$Z^{m}$  is invariant under scale transformations, but  $ Y^{m\mu}$ transforms as $ \rho^{\mu}$.
The scale-covariant derivaties for these  w-s densities  are defined  by the expressions  
\begin{eqnarray}\label{codZ,Y}
D_{\alpha}Z^{m}=\partial_{\alpha}Z^{m},  \,  \, \, \,  
 D_{\alpha}Y^{m \mu} = \partial_{\alpha}Y^{m\mu} - a_{\alpha} Y^{m \mu}.
\end{eqnarray}
Then the derivatives covariant under diff-s and scaling take the form 
\begin{eqnarray}\label{tocdZ,Y}
{\bf D}_{\alpha}Z^{m}=\partial_{\alpha}Z^{m} -  
\frac{1}{2}\Gamma^{\rho}{}_{\alpha\rho}Z^{m}, \nonumber 
\\  
{\bf D}_{\alpha}Y^{m\mu}= \partial_{\alpha}Y^{m\mu}  
+ \Gamma^{\mu}{}_{\lambda\alpha} Y^{m \lambda}  
- \Gamma^{\rho}{}_{\alpha\rho} Y^{m\mu} -  a_{\alpha}Y^{m\mu}  
 \end{eqnarray}
 After summing over the pair of w-s indices in Eq.(\ref{tocdZ,Y}), we obtain
\begin{eqnarray}\label{doptotcovd}
{\bf D}_{\mu}Y^{m\mu}=   
D_{\mu}Y^{m\mu}    \nonumber
\end{eqnarray}

Use of  $Z^{m}$ and  $\Sigma^{m \nu}$ permits to present  the Lagrangian 
density $\mathcal{L}$ for the action (\ref{actn}) as the scale-invariant w-s scalar density 
 $\mathcal{L}, \,  \omega_{\mathcal{L}}=1 $ 
\begin{eqnarray}\label{lagr}
\mathcal{L} =\frac{1}{2}Z^{m}Z_{m} + \frac{i}{2} \Sigma^{m}\rho^{\nu}D_{\nu} \Sigma_{m}, 
\end{eqnarray} 
 where we take into account the relation  $\Sigma^{m}\Sigma_{m} \equiv 0$.   
The fermionic kinetic term is linear in the velocities $\dot\Sigma^{m}$ that gives rise the well-known  fermionic constraint in the Hamiltonian formalism.  Dirac's approach explained how to deal with such cases.

The Euhler-Lagrange equations  following from  (\ref{actn})  in the fermionic sector have the form  
\begin{eqnarray} 
 &\delta\Sigma_{m}  \rightarrow&   Z^{m}\chi  -  \rho^{\nu} \partial_{\nu}\Sigma ^{m}
- \frac{1}{2}\partial_{\nu}\rho^{\nu} \Sigma ^{m}=0  \label{sigm},  \\
&\delta\chi \rightarrow&     Z_{m}\Sigma^{m} =0     \, \  \rightarrow  \, \,   Y^{m\mu}\Sigma_{m}= 0. \label{chi}
\end{eqnarray}
 Eq.(\ref{sigm}) multiplied by $\Sigma_{m}$  reveals the  vanishing of the fermionic  kinetic term
\begin{eqnarray}\label{nulkt}
\frac{i}{2}\Sigma_{m}\rho^{\mu}\partial_{\mu}\Sigma ^{m}=0
\end{eqnarray}
provided by the constraint  (\ref{chi}). 
 The  Euhler-Lagrange equations in  the bosonic sector are
\begin{eqnarray} 
&\delta X_{m} \rightarrow& D_{\mu}Y^{m\mu} =0,  \label{X}  \\
&\delta \rho^{\mu}  \rightarrow&   Z_{m}D_{\mu}X^{m} +\frac{i}{2}\Sigma_{m}\partial_{\mu}\Sigma ^{m}=0.\label{rho} 
\end{eqnarray}
After multiplication of (\ref{rho}) by $\rho^{\mu}$ and using Eq.(\ref{nulkt}) we find  
\begin{eqnarray}\label{Yort} 
Z^{m } \rho^{\mu} D_{\mu}X_{m}\equiv Y^{m\mu}D_{\mu} X_{m}= 0.
\end{eqnarray}
Eq.(\ref{Yort}) together with (\ref{X}) yield the zero scale-covariant divergence
\begin{eqnarray}\label{deort}
D_{\mu} (Y^{m\mu}X_{m}) =
\partial_{\mu} (Y^{m\mu}X_{m}) =0,
\end{eqnarray}
 where we used that both $ Z^{m }$ and $\rho^{\mu} X^{m }$ are scale invariants. Relation  (\ref{deort}) yields the conservation law of  the w-s  dilatation current $j_{dil.}^{\mu}(\xi):= Y^{m\mu}X_{m}$
 \begin{eqnarray}\label{dcur}
\partial_{\mu}j_{dil.}^{\mu}=\partial_{\mu} (Y^{m\mu}X_{m})=0.
\end{eqnarray}
Note the absence of a contribution  to the current $j_{dil.}^{\mu}(\xi)$ from kinetic terms  of   fermionic superpartners, which is equal to $\frac{i}{4}\Sigma ^{m}\rho^{\mu} \Sigma _{m}$, but vanishes due to the identity $\Sigma^{2}\equiv0$.

Next, using the definiton  (\ref{Z,Y}) of $Z^{m}$ we find  the scale-invariant constraint  $Z^{2}=0$
\begin{eqnarray}\label{Z2}
Z_{m}Z^{m}\equiv Z_{m}\rho^{\nu}D_{\nu}X^{m} + i\chi Z_{m}\Sigma^{m}=0
\end{eqnarray}
enforced by  (\ref{chi}) and  (\ref{Yort}).  Eqs.(\ref{Z2}), (\ref{nulkt}) imply the vanishing of the Lagrange density (\ref{lagr}) 
\begin{eqnarray} \label{lagr0}  
\mathcal{L}|_{(\ref{nulkt}), (\ref{Z2})}=0.
\end{eqnarray}
 The bosonic  Eqs.(\ref{X}-\ref{rho})  are complemented by the variational equation  for $a_{\mu}$  
\begin{eqnarray} \label{not}  
 &\delta a_{\mu}  \rightarrow&  Y^{m\mu} X_{m}= 0  \, \,  \rightarrow \, \,
  Z^{m} X_{m}= 0.
\end{eqnarray}   

To clarify the above Lagrangian constraints we note that  $Y^{m\mu}$ is the partial derivative
\begin{eqnarray} \label{dder} 
Y^{m\mu}=\frac{\partial \mathcal{L}}{\partial\partial_{\mu}X_{m}}
\end{eqnarray}  
with the  time component $Y^{m\tau}$ defining the momentum density
 $\mathcal{P}^{m}$ in the Hamiltonian approach 
 \begin{eqnarray} \label{mome}
\mathcal{P}_{m}:=Y_{m}^{\tau}=\frac{\partial \mathcal{L}}{\partial\dot{ X}^{m}}
  \, \  \rightarrow  \, \, \mathcal{P}_{m}= \rho^{\tau}Z_{m}.
 \end{eqnarray}  
The square of $\mathcal{P}_{m}$ yields the $\tau$-shift generator known from string models 
\begin{eqnarray} \label{squarmo} 
\mathcal{P}_{m}\mathcal{P}^{m}=Y_{m}^{\tau}Y^{m\tau}
=(\rho^{\tau})^{2} Z_{m}Z^{m}=0. 
\end{eqnarray}
  Comparison of  (\ref{Z2}) and (\ref{squarmo})  shows that the latter 
yields a Lagrangian image of the constraint $\mathcal{P}^{2}=0$ characterizing  tensionless spinning string.  

Eq.(\ref{not}) produces a Lagrangian image of the Hamiltonian constraint associated with local scale transformations. This constraint  complements the Lagrange constraint (\ref{chi}) associated with the w-s supersymmetry by  substituting $X^{m}$ instead of   $\Sigma^{m}$.
However, it should be taken into account that  $Y^{m\mu}X_{m}$ is a w-s vector density unlike $Y^{m\mu}\Sigma_{m} $ which has the weight $w_{Y\Sigma}=5/4$. 

The constraint  (\ref{rho}) creates a Lagrangian image of the  Hamiltonian constraint 
 \begin{eqnarray} \label{shift}
l:=\mathcal{P}_{m}\partial_{\sigma}X^{m} + \frac{i}{2}\rho^{\tau}\Sigma_{m}
\partial_{\sigma}\Sigma^{m}=0
\end{eqnarray}
generating shifts of the w-s coordite $\sigma$.
This follows after multiplication of (\ref{rho}) by $\rho^{\tau}$, choosing 
$\mu=\sigma$ and using the equation of motion (\ref{not}) together with its consequence 
\begin{eqnarray}\label{YderX}
Y^{m\mu}D_{\nu}X_{m}=Y^{m\mu}\partial_{\nu}X_{m}.
\end{eqnarray}
Moreover, multiplication of  (\ref{rho}) by $\rho^{\nu}$ creates the tensor density $T^{\nu}_{\mu}$ with $w_{T}=1$
\begin{eqnarray}\label{E-Mt}
 T^{\nu}_{\mu}:=Y^{m\nu}D_{\mu}X_{m} +\frac{i}{2}\rho^{\nu}\Sigma_{m}\partial_{\mu}\Sigma ^{m}=0
\end{eqnarray}
which vanishes after using the equation of motion for $\rho^{\mu}$.  
In addition, $T^{\nu}_{\mu}$ is traceless due to (\ref{nulkt}), (\ref{Yort}).
Taking into account the constraint (\ref{not}) permits to change $D_{\mu}X^{m}$ with  $\partial_{\mu}X^{m}$ in (\ref{E-Mt})
\begin{eqnarray}\label{E-Mtno}
 T^{\nu}_{\mu}|_{(\ref{not})}:=Y^{m\nu}\partial_{\mu}X_{m} +\frac{i}{2}\rho^{\nu}\Sigma_{m}\partial_{\mu}\Sigma ^{m}=0.
\end{eqnarray}
All that hints to consider $T^{\nu}_{\mu}$ as the density of the energy momentum tensor modulo possible terms disappearing on the mass shell. 

To verify the above conjecture, we recall that under transition 
from local symmetries to global ones, such as translational symmetry, in particular, the Noether identities result in the emergence of conserved currents for matter fields. Applying this criterion to two-dimensional translations  
we obtain the following Noether current  
\begin{eqnarray}\label{E-Mtok}
 T^{\mu}_{\nu}=Y^{\mu}_{m}\partial_{\nu}X^{m} +\frac{i}{2}\rho^{\mu}\Sigma_{m}\partial_{\nu}\Sigma^{m} -  \delta^{\mu}_{\nu}\mathcal{L}
\end{eqnarray}
which coincides with (\ref{E-Mtno}) due to the condition (\ref{lagr0}). 
So, we accept that  the density of the energy-momentum tensor of the tensionless spinning string is defined  by (\ref{E-Mtok}). The latter coincides with (\ref{E-Mtno}) if  (\ref{lagr0}) is taken into account.   
 Thus, $T^{\mu}_{\nu}$ defined by (\ref{E-Mtok}),  vanishes on the equations of motion as well as the hamiltonian densiy $\mathcal{H}$.

\section{Energy-momentum tensor density}

 The conjecture that $T^{\mu}_{\nu}$  (\ref{E-Mtok}) is the  density of the energy-momentum tensor means that it is equal to zero.  As a consequence its  trace $T^{\mu}_{\mu}$ is also  zero 
\begin{eqnarray}\label{traceT}
T^{\mu}_{\mu}=Y^{\mu}_{m}\partial_{\mu}X^{m} +\frac{i}{2}\rho^{\mu}\Sigma_{m}\partial_{\mu}\Sigma^{m} -  2\mathcal{L} =0
\end{eqnarray} 
 as seen from Eqs.(\ref {E-Mtno}) and (\ref{lagr0}). 
The latter shows that the w-s vector $P^{\mu}$ vanishes
\begin{eqnarray}\label{Pconsr}
P^{\mu}=\int d\sigma ( Y_{m}^{\mu}\partial_{\tau}X^{m} +  \frac{i}{2}\rho^{\mu}\Sigma_{m}\partial_{\tau}\Sigma^{m}  -  \delta^{\mu}_{\tau}\mathcal{L})=0.
\end{eqnarray} 
 Since the time component $P^{\tau}$ of  $P^{\mu}$ defines the canonical Hamiltonian H, one can see that H=0 (in the weak sense), as it must be for systems with the first class constraints.
Preservation in time of the constraint H=0 means that the covariant derivative $T^{\mu}_{\nu}$ must be zero
\begin{eqnarray}\label{DT}
{\bf D}_{\mu}T^{\mu}_{\nu}=0.
\end{eqnarray} 
This covariant condition is equivalent to preserving the first order constraints required in the  Dirac constraint theory.
Therefore, to prove the conjecture that $T^{\mu}_{\nu}$ (\ref{E-Mtok}) is the energy-momentum tensor density, we need to prove Eq.(\ref{DT}).
Covariant differentiation of (\ref{E-Mtok}) gives 
\begin{eqnarray}\label{cvdE-M}
 {\bf D}_{\mu}T^{\mu}_{\nu}=
{\bf D}_{\mu}(Y^{\mu}_{m}\partial_{\nu}X^{m}) +
\frac{i}{2}{\bf D}_{\mu}(\rho^{\mu}\Sigma_{m}\partial_{\nu}\Sigma^{m}) 
- {\bf D}_{\nu}\mathcal{L} \nonumber  \\
= \{ {\bf D}_{\mu}(Y^{\mu}_{m}\partial_{\nu}X^{m}) - \frac{i}{2}{\bf D}_{\nu}(Z^{m}Z_{m}) \}
+\frac{i}{2}\{{\bf D}_{\mu}( \rho^{\mu}\Sigma_{m}\partial_{\nu}\Sigma^{m})
  -    {\bf D}_{\nu}(\Sigma_{m}\rho^{\mu}\partial_{\mu}\Sigma^{m}) \},
 \end{eqnarray}
where the representation (\ref{lagr}) of  $\mathcal{L}$ is used. The first term in Eq.(\ref{cvdE-M}) represents the covariant contribution of the bosonic fields into ${\bf D}_{\mu}T^{\mu}_{\nu}$ equal to 
\begin{eqnarray}\label{boson}
{\bf D}_{\mu}T^{\mu}_{\nu}|_{boson} 
=- Z_{m} D_{\mu} X^{m} [\partial_{\nu}\rho^{\mu} + \Gamma^{\mu}{}_{\nu\lambda} 
\rho ^{\lambda}] - i\chi Z_{m} \partial_{\nu}\Sigma^{m}
\end{eqnarray}
as follows from (\ref{tocdZ,Y}). The second term in (\ref{cvdE-M}) gives a similar contribution from fermions
\begin{eqnarray}\label{fermion}
{\bf D}_{\mu}T^{\mu}_{\nu}|_{fermion} =
-\frac{i}{2}\Sigma_{m} D_{\mu} \Sigma^{m} [\partial_{\nu}\rho^{\mu} + \Gamma^{\mu}{}_{\nu\lambda}\rho ^{\lambda}] + i\chi Z_{m} \partial_{\nu}\Sigma^{m}.
\end{eqnarray}
Then summing the contributions of the bosonic and fermionic  fields we obtain 
\begin{eqnarray}\label{tcvdE-M}
 {\bf D}_{\mu}T^{\mu}_{\nu}:= {\bf D}_{\mu}T^{\mu}_{\nu}|_{boson} +{\bf D}_{\mu}T^{\mu}_{\nu}|_{fermion}= -(Z_{m} D_{\mu} X^{m} + \frac{i}{2}\Sigma_{m} D_{\mu} \Sigma^{m})[\partial_{\nu}\rho^{\mu} + \Gamma^{\mu}{}_{\nu\lambda}\rho ^{\lambda}].
\end{eqnarray}
Here the first factor  in (\ref{tcvdE-M})  vanishes according to the equation of motion (\ref{rho}).

This proves our conjecture that the density of the energy-momentum tensor of N=1 spinning string  is defined by  relation (\ref{E-Mtok}) and its covariant divergence  (\ref{DT})  is equal to zero
$${\bf D}_{\mu}T^{\mu}_{\nu}=0.$$
It is also interesting to note that covariant conservation of $T^{\mu}_{\nu}$ is achieved by exact cancellation between its bosonic and fermionic contributions at pseudoclassical level.

\section{Summary}

In this paper  we continue study of dynamics and w-s geometry of tensionless spinning string with the action invariant under reparametrizions, both local supersymmetry and dilatations.  This model realizes the limit of zero tension in the well-known model of tensile spinning string treated as a model of two-dimensional supergravity interacting with matter.   Geometries of worldsheets swept by tensionless and tensionfull spinning strings are compared.  It is shown that for the  tensionless string the full w-s  torsion tensor is  defined  incompletely. Its contribution is represented  by the covariant trace $S_{\mu}$, whereas for the tensile string all the tensor componets are defined by the bilinear covariant $\sim i{\bar\chi}_{\mu}\gamma_{5}\gamma^{\rho}\chi_{\rho}$ in the w-s gravitino $\chi_{\mu}$.  This  is due to loss of both spinorial structure of nilpotent density (\ref{varchim}) and some  components of the affine connection 
$\Gamma^{\alpha}{}_{\mu}{}_{\beta}$.  The  remaining components $\Gamma_{\mu}(\xi)$ in covariant derivatives of tensor densities in $S$ create
 a protective mechanism against elimination of torsion from the action of tensionless spinning strings. Further we assume that  in string picture the Grassman scalar density $\Sigma^{m}(\xi)$,  which is fermionic partner of $X^{m}(\xi)$,  can be interpreted as an order parameter describing relativistically covariant collectivization of the space-time spin variables of interacting massless spinning particles. 

The  correspondence  between relativistically invariant physics in target multidimensional (super)spaces and on hypersurfaces embedded in them
may be useful for  the choice of true physical vacuum and other  applications in relativistic systems containing spin degrees of freedom.  In particular, it  seems  interesting  to estimate  the role of tensionless extended objects in search for  solution of  the landscape problem in string theory,  where  a huge number of vacuum states  $\sim 10^{500}$ is predicted. 
The generalized Euhler-Lagrange equations and  the Lagrangian consraints  are 
derived and expressed  in terms of covariant blocks constructed from w-s fields, densities and their extended  covariant derivatives.  Connection of these constraints with the Hamiltonian ones is discussed.  Using these equations and constraints,  the covariant density  of energy-momentum tensor for the tensionless spinning string is derived. We  prove vanishing of this density and its covariant divergence.  This result arises from mutual cancellation of the bosonic and fermionic contributions.  
We hope that the above discussed  peculiarities of models of tensionless spinning strings extended  by addition of dilatation symmetry may shed new light on hidden relations between  (super)gauge theories and (super)gravity. 

 \section{Asknowledgments}

The author is grateful to NORDITA for warm hospitality and support. I also wish to thank D. Uvarov and  H. von Zur-M\"{u}hlen for useful discussions.

\end{document}